\documentclass[a4paper,11pt,nohyper]{JHEP3}
\usepackage{amsmath,latexsym,amssymb,slashed}
\usepackage{graphicx}
\usepackage{caption}

\usepackage[latin1]{inputenc}
\usepackage[T1]{fontenc}
\usepackage{subcaption}

\graphicspath{{./Figures/}}
\newcommand\beal{\begin{align}}

\newcommand{\eq}[1]{\begin{equation}#1\end{equation}}
\newcommand{\spl}[1]{\begin{split}#1\end{split}}

\newcommand{\arXividhepth}[1]{\href{http://arxiv.org/abs/#1}arXiv:{\tt #1} [hep-th]}

\newcommand{\mcal}{\mathcal{M}}

\newcommand{\ncal}{\mathcal{N}}

\def\d{\text{d}}

\def\slashchar#1{\setbox0=\hbox{$#1$}           
\dimen0=\wd0                                 
\setbox1=\hbox{/} \dimen1=\wd1               
\ifdim\dimen0>\dimen1                        
\rlap{\hbox to \dimen0{\hfil/\hfil}}      
#1                                        
\else                                        
\rlap{\hbox to \dimen1{\hfil$#1$\hfil}}   
/                                         
\fi}

\title{On the spin-2 Kaluza-Klein spectrum of  AdS\boldmath$_4\times S^2(\mathcal{B}_4)$\unboldmath}

\author{Jean-Marc Richard${}^{\clubsuit}$, Robin Terrisse${}^{\clubsuit\diamondsuit}$ and Dimitrios Tsimpis${}^{\clubsuit}$
  \\

  \begin{itemize}
  
\item Universit\'{e} Claude Bernard (Lyon 1)\\
UMR 5822, CNRS/IN2P3, Institut de Physique Nucl\'{e}aire de Lyon\\ 
4 rue Enrico Fermi,  
F-69622 Villeurbanne Cedex, France

\item  Universit\'e de Lyon,  ENS de Lyon,\\
46 all\'ee d'Italie, F-69364 Lyon CEDEX 07, France 
  \end{itemize}

\bigskip
 E-mail:\\
\email{j-m.richard@ipnl.in2p3.fr}, \email{robin.terrisse@ens-lyon.fr} \&\\\email{tsimpis@ipnl.in2p3.fr} }

\abstract{
We perform a numerical study of the four-dimensional spin-2 Kaluza-Klein spectrum of supersymmetric AdS$_4\times S^2(\mathcal{B}_4)$ vacua 
and show that they do not exhibit scale separation. 
Our methods are generally applicable to similar problems where 
the compactification geometry is not known analytically, hence an analytic treatment of the spectrum of Kaluza-Klein masses is not 
available.}

\begin{document}
\setcounter{footnote}{0}
\renewcommand{\thefootnote}{\arabic{footnote}}
\setcounter{section}{0}
\section{Introduction}

\label{introduction}

All the supersymmetric, pure-flux (i.e., without external sources such as orientifolds) AdS solutions that have been constructed since the early days of supergravity suffer from the so-called problem of scale separation: supersymmetric backgrounds of the form AdS$\times\mcal$ (where the product is 
topologically direct but may be metrically warped) have the property that the radius of curvature of the  AdS space is of the order of the `radius' of the internal manifold $\mcal$.\footnote{For the purposes of the present paper 
we may think of the (radius)$^2$ of $\mcal$ as the inverse of the first nonzero 
eigenvalue of the scalar Laplacian on $\mcal$. Note however that the concept of a `radius' or a `diameter' of a space is mathematically more involved and sometimes inequivalent definitions exist in the mathematics literature.}

The absence of any counter-examples to this empirical observation may very well 
be the manifestation of some general underlying obstruction. However no such underlying no-go theorem has been proven to date. On the other hand, supersymmetric supergravity AdS vacua with scale separation, if they exist, would be highly desirable: if the ratio of radii of the AdS to the internal space could be tuned to be arbitrarily large, such vacua might be viable starting points for realistic compactifications. Moreover AdS vacua are in principle very well controlled to the extent that they can be defined nonperturbatively as quantum gravity theories via a dual conformal field theory.

The massive IIA $\ncal=2$ solutions of \cite{lt3} are of the form  AdS$_4\times S^2(\mathcal{B}_4)$, where $S^2(\mathcal{B}_4)$ is a two-sphere bundle over a four-dimensional K\"{a}hler-Einstein base $\mathcal{B}_4$. These solutions naively seem to have enough parameters to independently control the sizes of the external AdS$_4$ and the internal $S^2(\mathcal{B}_4)$ space. However \cite{Tsimpis:2012tu} observed an `atractor' behavior: 
although the warp factors of the solution that control the relative scales of the external and internal spaces can be chosen to have an arbitrarily large ratio at the north pole of the $S^2$, that ratio becomes necessarily of order one in a neighborhood the equator. 
Moreover it was shown that the asymptotic Kaluza-Klein (KK) spectrum of 
a ten-dimensional scalar in the background of \cite{lt3} is governed by a scale of the order of the AdS$_4$ radius. Nevertheless the results of \cite{Tsimpis:2012tu} were inconclusive as to the existence or absence of scale separation in the vacua of \cite{lt3}.

Given the potential phenomenological interest of AdS vacua with scale separation it is important to study this problem further. This is the purpose of the present paper. In order to settle the question of scale separation we examine directly the KK spectrum of four-dimensional spin-2 excitations. Absence of scale separation is equivalent to the existence of low-lying massive modes, that is modes with mass $M$ of the order of the inverse of the AdS radius of curvature $L$: $M L\sim\mathcal{O}(1)$. By a numerical analysis of the spectrum we have found that for any choice of the discrete data describing the solutions of \cite{lt3} there  
always exist low-lying massive graviton modes with masses: 
\eq{\label{bound}ML\lesssim 2.4~,}
thus conclusively showing the absence of scale separation in the vacua of \cite{lt3}.

The plan of the remainder of the paper is as follows. In section \ref{sec:2} we review 
the relevant properties of the supersymmetric AdS$_4\times S^2(\mathcal{B}_4)$ vacua. In section \ref{sec:s2} a subset of the four-dimensional spin-2 KK spectrum is mapped 
to the spectrum of eigenvalues of a SL problem. In section \ref{sec:s5} we explain the numerical method used in the analysis of the spectrum and we establish the bound (\ref{bound}) for the KK scale. We conclude in section \ref{sec:5}.

\section{AdS\boldmath$_4\times S^2(\mathcal{B}_4)$\unboldmath vacua}\label{sec:2}

For the convenience of the reader we summarize here the relevant properties of the solutions of \cite{lt3}; more details can be found in \cite{lt3,Tsimpis:2012tu}. These are  $\ncal=2$ warped AdS$_4\times \mcal_6$ type IIA supergravity backgrounds with metric (in the string frame) given by\footnote{The solutions of \cite{lt3} can be thought of as massive IIA deformations 
of the $\ncal=2$ IIA circle reductions of 
the M-theory  AdS$_4\times Y^{p,q}(\mathcal{B}_4)$ backgrounds of \cite{gm,ms2}, where $Y^{p,q}(\mathcal{B}_4)$ is a seven-dimensional Sasaki-Einstein manifold. 
The first such massive deformation was constructed in \cite{pz} for the $\ncal=2$ IIA circle reduction of the M-theory AdS$_4\times Y^{3,2}(\mathbb{CP}^2)$ background (the $Y^{3,2}(\mathbb{CP}^2)$ 
space is also referred to as $M^{1,1,1}$ in the physics literature).}
\eq{\label{1}
\d s^2_{10}=
e^{2A(\theta)}\d s^2(AdS_4)+L^{2}\d s^2(\mcal_6)
~.}
The metric $\d s^2(AdS_4)$ is that of a four-dimensional 
anti-de Sitter space of radius $L$, so that the scalar curvature is related to the radius through: $R=-12L^{-2}$. 
The metric of the internal space is given by
\eq{\label{3}
\d s^2(\mcal_6)=e^{2C(\theta)}\d s^2(\mathcal{B}_4)+e^{2A(\theta)}\big(
f^2(\theta)\d\theta^2+\sin^2\!\theta~\!(\d\psi+\mathcal{A})^2\big)
~,}
where
\eq{\label{4}
f(\theta):=\frac{1}{2-\sin^2\!\theta ~\!e^{2(A(\theta)-C(\theta))}}~.}
The coordinates $(\psi,\theta)$, with ranges 
$0\leq\psi< \pi$, $0\leq\theta\leq\pi$, parameterize a smooth $S^2$ fiber over $\mathcal{B}_4$; the coordinate $\psi$ parameterizes an $S^1$ fiber in the anticanonical bundle of $\mathcal{B}_4$. The $U(1)$ connection $\mathcal{A}$ on $\mathcal{B}_4$ is related to the K\"{a}hler form $J$ on $\mathcal{B}_4$ through
\eq{\label{6}
\d\mathcal{A}=-J
~,}
while the four-dimensional K\"{a}hler-Einstein metric of $\mathcal{B}_4$ is canonically normalized. 

The dependence of the functions  $A$, $C$ on the coordinate $\theta$ is given implicitly through the following system of two coupled first-order differential equations:
\eq{\spl{\label{7}
A'&=\frac{1}{2}\tan\theta\frac{1-\sin^2\!\theta ~\!e^{2(A-C)}}{2-\sin^2\!\theta ~\!e^{2(A-C)}}\\
C'&=\frac{1}{4}\sin(2\theta)\frac{
e^{2(A-C)}}{2-\sin^2\!\theta ~\!e^{2(A-C)}}
\frac{1+e^{8A}}{1+\cos^2\theta e^{8A}}
~,}}
where the primes denote differentiation with respect to $\theta$. 
The system (\ref{7})  has not been solved analytically to date. On general grounds,  for a given set $(A_0,C_0)$ of `initial conditions'
\eq{
A_0:=\left.A\right|_{\theta=0}~,~~~C_0:=\left.C\right|_{\theta=0}
~,}
we expect a unique solution at least in a neighborhood of $\theta=0$. 
On the other hand, by virtue of (\ref{3}), we expect that the parameters  $(A_0,C_0)$ should control the size of the external AdS$_4$ and the internal $S^2(\mathcal{B}_4)$ space. 
However, as observed in \cite{Tsimpis:2012tu}, this expectation is not entirely correct: although $(A_0,C_0)$ can be chosen to have an arbitrarily large ratio at the north pole of the $S^2$ ($\theta=0$), that ratio becomes necessarily of order one in a neighborhood the equator ($\theta=\pi/2$). 
Finally let us also note that upon imposing flux quantization the parameter space of the initial values $(A_0,C_0)$ becomes discretized \cite{Tsimpis:2012tu}.

\section{The spin-2 Kaluza-Klein spectrum}\label{sec:s2}

For the analysis of the KK spectrum of four-dimensional spin-2 fields (massive ``gravitons'') we will 
draw upon the results of \cite{Bachas:2011xa} where it was shown  (generalizing earlier work of \cite{Csaki:2000fc}) that the spin-2 excitations of any ten-dimensional background containing a $d$-dimensional factor with maximal symmetry (i.e., AdS$_d$, $\mathbb{R}^{1,d-1}$ or dS$_d$) obey the massless scalar ten-dimensional wave equation. In particular for supergravity backgrounds  of the form (\ref{1}) this result correlates the KK mass of four-dimensional gravitons to the eigenvalues of a modified Laplacian of $\mcal_6$. 

More specifically, let $\bar{g}_{\mu\nu}$ be the metric of the AdS$_4$ space  appearing in (\ref{1}) and consider four-dimensional metric perturbations of 
the form
\eq{\label{metricpert}\d s^2_{10}=
e^{2A}\left(\bar{g}_{\mu\nu}+h_{\mu\nu}\right)\d x^{\mu}\d x^{\nu}+L^{2}\d s^2(\mcal_6)
~,}
where $x^{\mu}$, $\mu=0,\dots, 3$, are coordinates of AdS$_4$. 
Furthermore let us expand the perturbation $h_{\mu\nu}$ as follows:
\eq{\label{f}
h_{\mu\nu}(x,y)=\sum_n h^{(n)}_{\mu\nu}(x)g_n(y)
~,}
where $y$ denotes the coordinates of $\mcal_6$; the $g_n(y)$'s are orthonormal weighted eigenfunctions of a modified Laplacian\footnote{Equation (\ref{int}) is equivalent to (2.20) of \cite{Bachas:2011xa} where $m$ and $\psi$ of that reference are identified 
respectively with $LM_n$ and $g_n$ here. 
Equation (\ref{pf}) of the present paper then reduces to (2.4) of \cite{Bachas:2011xa} provided 
we specialize to an AdS space of unit radius, $L=1$; this corresponds to setting $k=-1$ in \cite{Bachas:2011xa}. Note also that (5.3) of \cite{Tsimpis:2012tu} 
reduces to (\ref{int}) of the present paper upon setting $M=0$ in that reference and identifying $\lambda_n$ there  with $L^2M_n^2$ here.} of  $\mcal_6$ to eigenvalues $L^2M_n^2$:
\eq{\label{int}
-\frac{1}{\sqrt{g_6}}\partial_p
\left(\sqrt{g_6}g^{pq}e^{4A}\partial_qg_n(y)\right)=L^2M_n^2e^{2A}g_n(y)
~,}
with $g_{pq}$ the metric of  $\mcal_6$ and $g_6$ its determinant; $h^{(n)}_{\mu\nu}(x)$ in the expansion (\ref{f}) is assumed to be 
transverse and traceless and to obey the 
Pauli-Fierz equations for a massive spin-2 particle of mass $M_n$ in an 
AdS$_4$  space of radius of curvature $L$ (see, e.g., \cite{Duff:1986hr}):
\eq{\label{pf}\left(\bar{\nabla}^2
+2 L^{-2}-M_n^2
\right)h^{(n)}_{\mu\nu}(x)
~;~~~\bar{\nabla}^{\mu}h^{(n)}_{\mu\nu}(x)=0~;~~~
\bar{g}^{\mu\nu}h^{(n)}_{\mu\nu}(x)=0
~,}
where $\bar{\nabla}$ is the covariant derivative with respect to the Christoffel connection of $\bar{g}_{\mu\nu}$. It then follows from the analysis of \cite{Bachas:2011xa} that the metric (\ref{metricpert}) obeys the ten-dimensional linearized Einstein equations, independently of the form of the energy-momentum tensor for the matter fields (i.e., all fields of the theory other than the metric).

Normalizable spin-2 excitations correspond to eigenmodes $g_n$ of the modified Laplacian (\ref{int}) for which 
\eq{\label{norm}\int_{\mcal_6}\d^6y \sqrt{g_6}e^{2A}|g_n|^2<\infty~.}
Moreover \cite{Bachas:2011xa} shows that 
\eq{\label{mass}M_n^2\geq 0~,}
with the lower bound saturated, $M_0=0$, if and only if the corresponding 
eigenmode is constant: $g_0=\mathrm{const}$. In order to derive the bound above one simply multiplies (\ref{int}) by $g_n$ and integrates by parts, 
assuming that the integral of the total derivative does not pick up any contributions from singularities:
\eq{\label{intcond}\int_{\mcal_6}\d^6y\partial_p
\left(\sqrt{g_6}g^{pq}e^{4A}g_n\partial_qg_n\right)=0~.}

\subsection{Sturm-Liouville}\label{sec:sl}

To show the absence of scale separation it will suffice to consider the special case for which the eigenmodes only depend on the azimuthal  
angle of the $S^2$ fiber: $g_n=g_n(\theta)$. As shown in \cite{Tsimpis:2012tu}, equation (\ref{int}) reduces in that case 
to the following second-order ordinary differential 
equation (where a prime denotes 
differentiation with respect to $\theta$):
\eq{
\big(p~\!g_n'\big)'
+L^2M_n^2 ~\!q~\!g_n=0 ~,\label{Sturm}}
where:
\eq{\label{coeffs}
p(\theta):=\frac{e^{4(A+C)}}{f}~\!\sin\theta~,~~~
q(\theta):= e^{4(A+C)}f\sin\theta ~,
}
$L$ is the AdS$_4$ radius and $f(\theta)$ was given in (\ref{4}).  
Equation (\ref{Sturm}) is a singular Sturm-Liouville (SL) problem, since $p$ vanishes linearly at the endpoints $0$, $\pi$ of the $\theta$-interval, however it was shown in \cite{Tsimpis:2012tu} that it can be mapped to a {\it regular} SL problem by means of the transformation $\tilde{g}_n:=g_n/u$ where the function $u$ is given by:
\eq{\label{udef}
u=1-\log(\sin\theta)~.
}
Indeed $\tilde{g}_n$ is a solution of the following regular SL problem to the same eigenvalue $L^2M_n^2$:
\eq{
\big(\tilde{p}~\!\tilde{g}_n'\big)'
+(L^2M_n^2 \tilde{q}-\tilde{r})\tilde{g}_n=0
~,\label{Sturmreg}}
where
\eq{\label{trnsf}
\tilde{p}:=pu^2~,~~~ \tilde{q}:=qu^2~,~~~ \tilde{r}:=-u\big(pu'\big)'~.}
In order for the SL problem to be well-posed one needs to specify appropriate boundary conditions. Crucially, as we will see in section \ref{sec:s5} below, conditions 
(\ref{norm}),(\ref{intcond}) translate in the present case to the boundary conditions:
\eq{\label{sep}
\left.(\tilde{p}~\!\tilde{g}_n')\right|_{\theta=0}=
\left.(\tilde{p}~\!\tilde{g}_n')\right|_{\theta=\pi}=0
~.}
These are called `separated' boundary conditions \cite{zbook}
in the mathematics literature and they are admissible boundary conditions for the regular SL problem. 
Hence the subset of the four-dimensional spin-2 KK spectrum governed by (\ref{Sturm}) corresponds to the eigenvalues of the regular SL problem (\ref{Sturmreg})-(\ref{sep}).

It is well-known that the regular SL problem defined in (\ref{Sturmreg})-(\ref{sep}) has solutions $\tilde{g}_n(\theta)$ only for certain values of $M^2_n$. Specifically for separated boundary conditions as 
is the case here the eigenvalues can be indexed by $n\in\mathbb{N}$; they are bounded below and can be ordered to satisfy (\cite{zbook}, p.72):
\eq{\label{eigen}
0=M^2_0< M^2_1< M^2_2<\dots;~~M^2_n\rightarrow
+\infty~,~~ \mathrm{as}~~n\rightarrow\infty
~,}
where in the first equality above we took (\ref{mass}) into account. 
Moreover, the eigenfunction $g_n(\theta)$ corresponding to the eigenvalue $M^2_n$ 
has exactly $n$ nodes (zeros) in the interior of the interval $(0,\pi)$.


\section{Numerical Analysis}
\label{sec:s5}

\begin{figure}
\begin{subfigure}{.5\textwidth}
  \centering
  \includegraphics[width=\linewidth]{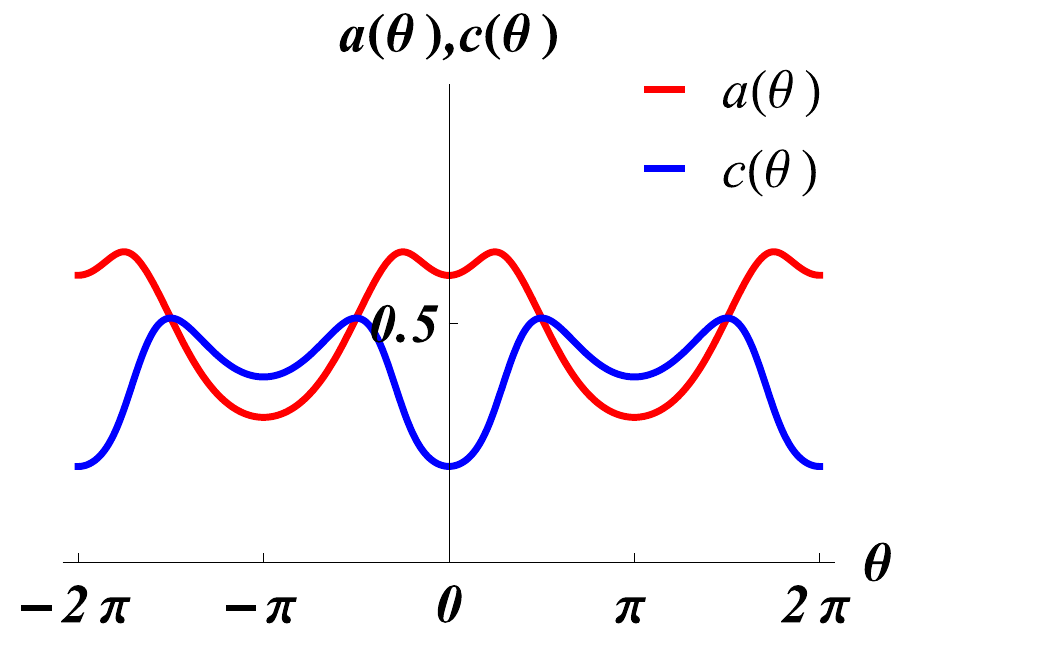}
  \caption{\label{Solutionadmissible} Admissible solution.}
\end{subfigure}
\begin{subfigure}{.5\textwidth}
  \centering
  \includegraphics[width=\linewidth]{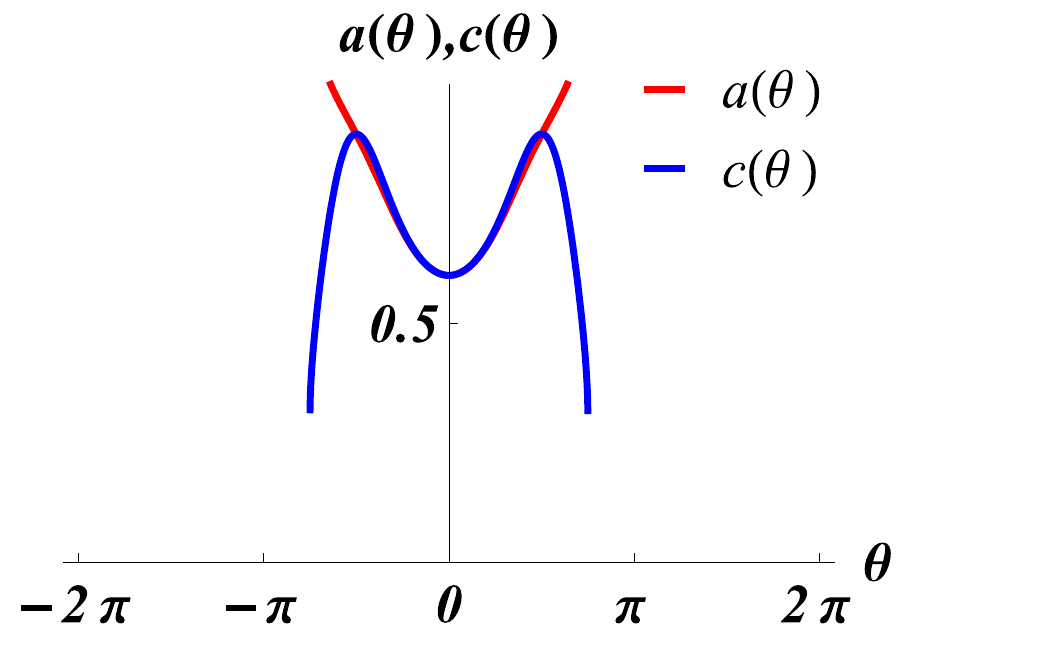}
  \caption{\label{Solutionnonadmissible}Inadmissible solution.}
\end{subfigure}
\caption{\label{exsol}Examples of solutions of \eqref{7} for two different sets of IC.}
\end{figure}

As a first step we will concentrate in section \ref{sec:da} on solving numerically the 
 system (\ref{7}) of two coupled first-order differential equations. 
To simplify the notation let us introduce two new variables: $a:=e^{2A}$ and $c:=e^{2C}$. 
We will be looking for solutions that are well-defined on $\theta \in \left[0,\pi \right]$; in particular $a(\theta)$, $c(\theta)$ should be bounded and non-negative by virtue of their definition. This is not automatically guaranteed to be the case, as can be seen for example in figure \ref{exsol}. The requirement that the solutions should be well-defined will impose constrains on the admissible domain ($D$) of our initial conditions (IC). In section \ref{sec:da} we will determine the space $D$ of all admissible IC.

Once $a$ and $c$ have been determined as functions of the IC, we will turn in section \ref{sec:vp} to the numerical analysis of the eigenvalue equation  (\ref{Sturm}). We will specify the boundary conditions (BC) at $\theta = 0$ and $\theta =\pi$ so that the corresponding eigenvalue equation (\ref{Sturmreg}), whose spectrum (\ref{eigen}) is the same as the spectrum of eigenvalues of (\ref{Sturm}), is a regular SL problem. 
We will then determine the first nonvanishing KK mass as a function of the IC, $M_1(\mathrm{IC})$ with IC $\in D$, and prove that it satisfies the bound (\ref{bound}) of the introduction.

\begin{figure}
\begin{subfigure}{.5\textwidth}
  \centering
  \includegraphics[width=\linewidth]{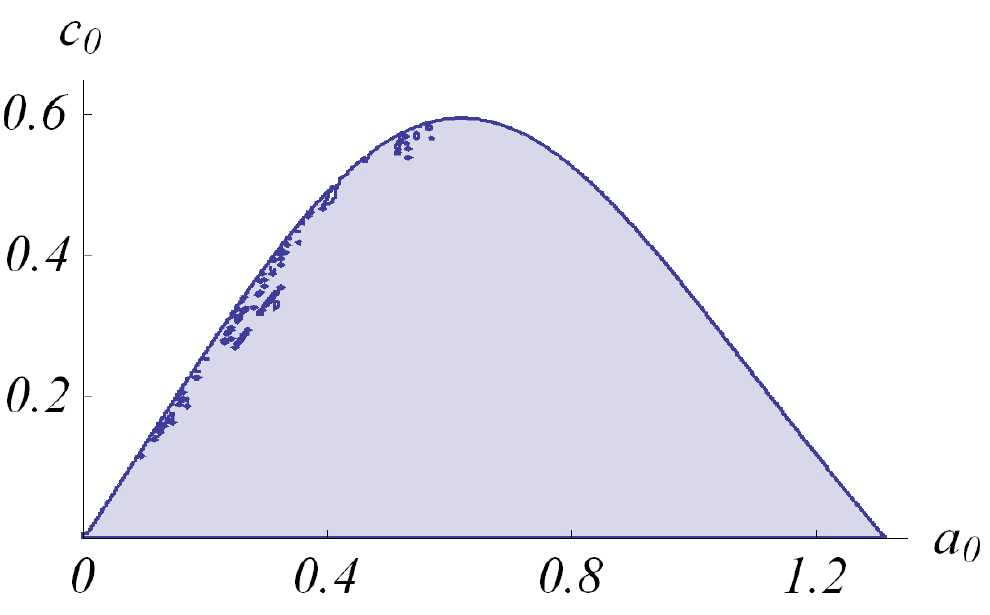}
  \caption{\label{Domaine1} Admissible domain in the $(a_0,c_0)$ plane.}

\end{subfigure}
\begin{subfigure}{.5\textwidth}
  \centering
  \includegraphics[width=\linewidth]{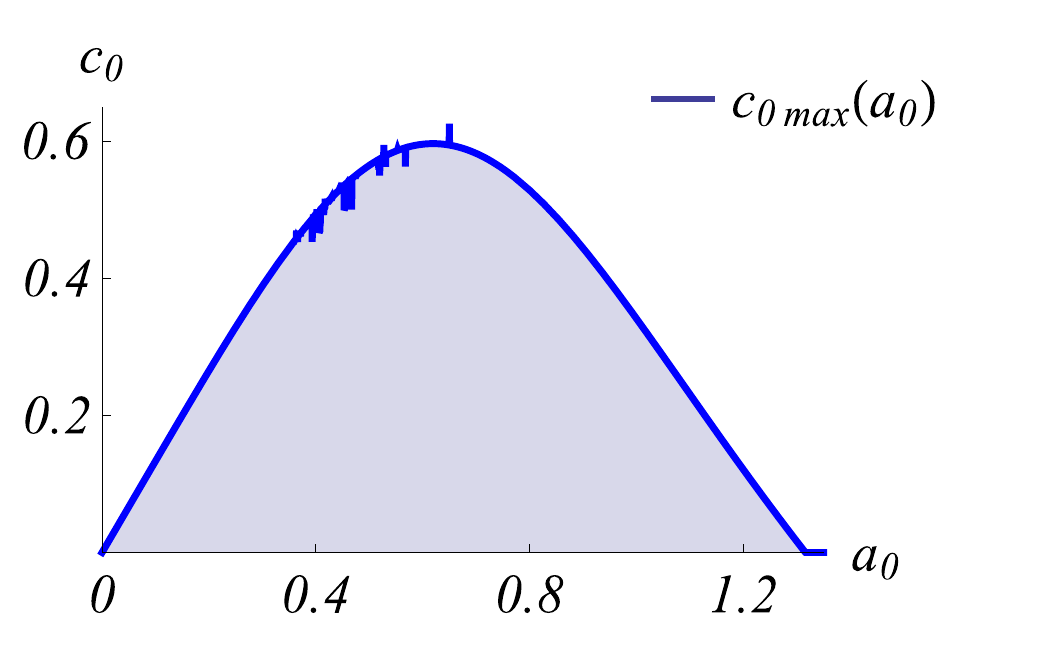}
  \caption{\label{c0max1} Plot of $c_{0,max}$ as a function of $a_0$.}

\end{subfigure}

\caption{\label{Domaines1} Determination of the admissible domain.}
\end{figure}

\subsection{Admissible domain}\label{sec:da}

Let us first consider imposing the IC $a=a_0$, $c=c_0$ at $\theta=0$. 
By virtue of their definition $a$, $c$ must be non-negative. We will thus look for the space $D$ of all IC such that $a(\theta)$, $c(\theta)$ are well-defined everywhere on $[0,\pi]$. 
It is then straightforward to plot the domain $D$ using Mathematica \cite{math} (figure \ref{Domaine1}). We may also plot the function $c_{0, max}$ defined as the maximum value of $c_0$ for a given $a_0$ (figure \ref{c0max1}).

\begin{figure}
\begin{subfigure}{.5\textwidth}
  \centering
  \includegraphics[width=\linewidth]{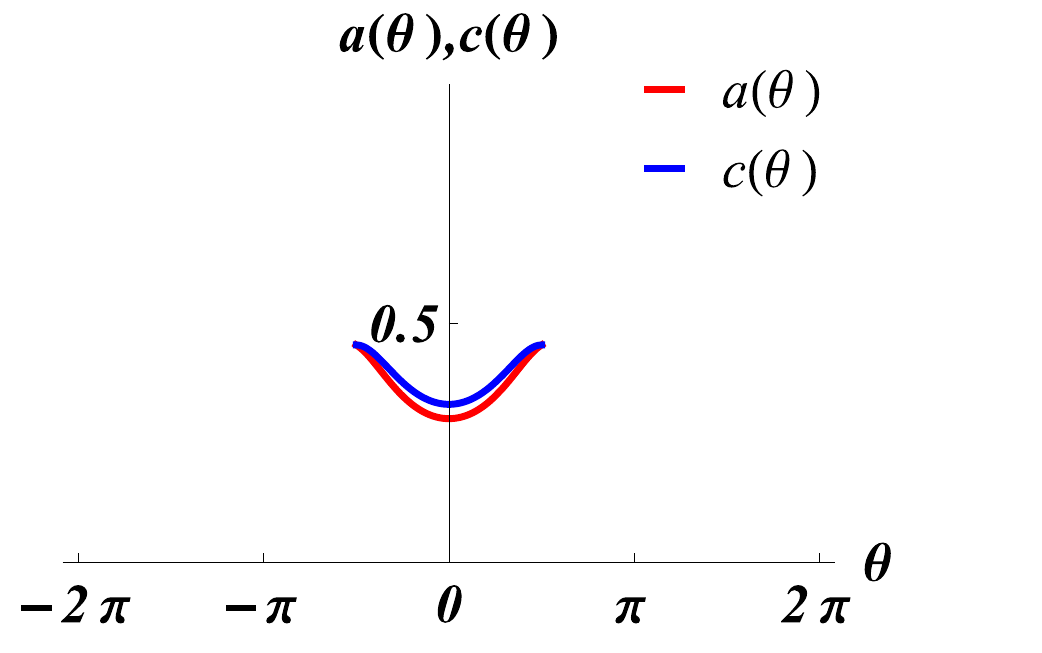}
  \caption{\label{Erreurnuma}Solution with automatic parameters}

\end{subfigure}
\begin{subfigure}{.5\textwidth}
  \centering
  \includegraphics[width=\linewidth]{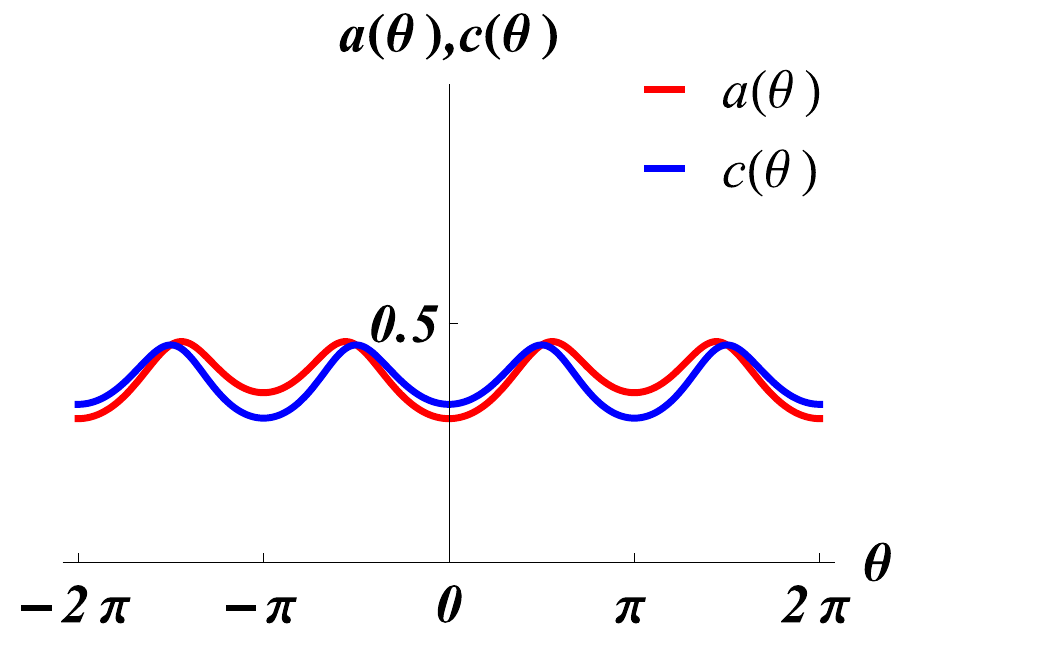}
  \caption{\label{Erreurnumb}Solution with different precision parameters.}

\end{subfigure}

\caption{\label{Erreurnum}Example of numerical instabilities in the solution of  equation \eqref{7}: with the right choice of parameters in the algorithm, a regular solution is obtained in an apparently divergent case.}
\end{figure}

\begin{figure}
\begin{subfigure}{.5\textwidth}
  \centering
\includegraphics[width=7.6cm,viewport=0 2 300 178,clip]{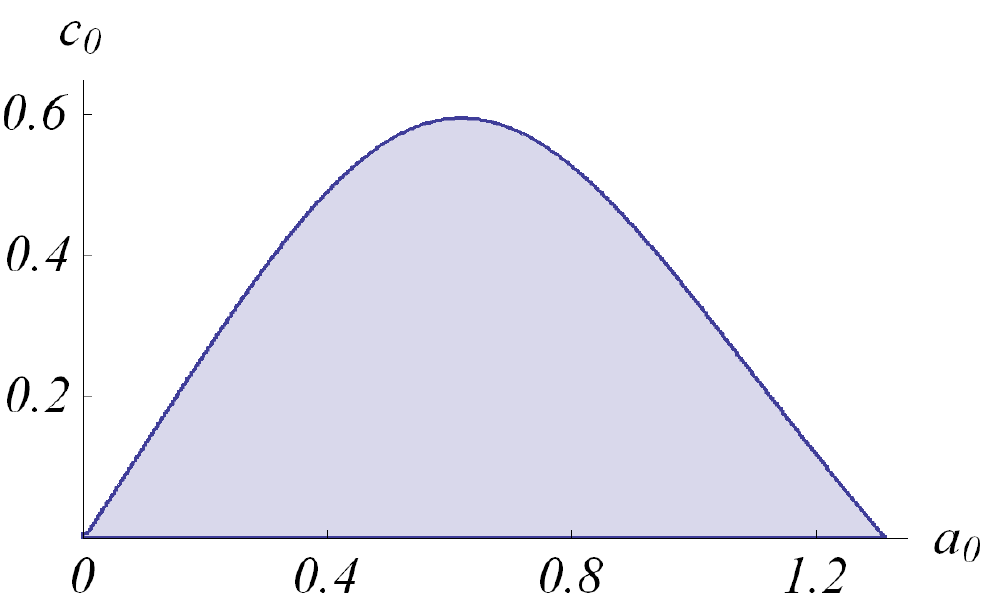}
  \caption{\label{Domaine2} Admissible domain in the $(a_0,c_0)$ plane.}

\end{subfigure}
\begin{subfigure}{.5\textwidth}
  \centering
 \scalebox{0.94}{ \includegraphics[width=7.8cm]{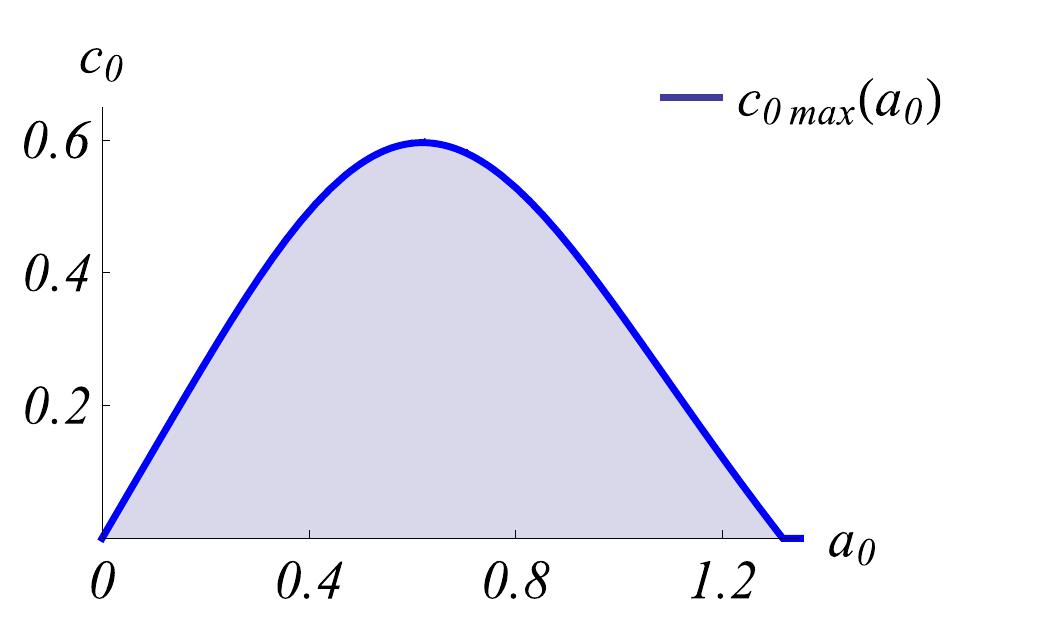}}
  \caption{\label{c0max2} Plot of $c_{0,max}$ as a function of $a_0$.}

\end{subfigure}

\caption{\label{Domaines2} Determination of the admissible domain using the BDF method: all apparent irregularities have disappeared.}
\end{figure}

In figure \ref{Domaines1} we notice certain irregularities in the admissible domain, which may signal the presence of numerical instabilities. It is therefore important to know whether the irregularities are artifacts of the numerical resolution. The plot of  $a$, $c$ at a point in $D$ corresponding to 
such irregularities  (figure \ref{Erreurnuma}) reveals that these problems come from the neiborhood of  $\theta=\frac{\pi}{2}$. In fact this can easily be understood as follows: at $\theta=\frac{\pi}{2}$ the derivative of  $a$ diverges unless $a(\frac{\pi}{2})=c(\frac{\pi}{2})$. The system is therefore very sensitive
at this point and susceptible to numerical instabilities. It is natural to suspect that this accounts for the irregularities in $D$. 
Indeed it may be seen that changing the precision parameters of the numerical resolution eliminates the problem (figure \ref{Erreurnumb}),  confirming the presence of numerical instabilities. 
To rectify this we have used a method of numerical resolution, implemented 
in Mathematica, 
which goes under the name "Backward Differentiation Formula" (BDF). 
This method is more robust and thus more appropriate for dealing 
with unstable systems and indeed eliminates all problems of 
numerical 
instabilities, leading to a regular domain (figure \ref{Domaines2}).


\begin{figure}
\begin{subfigure}{.5\textwidth}
  \centering
  \includegraphics[width=\linewidth]{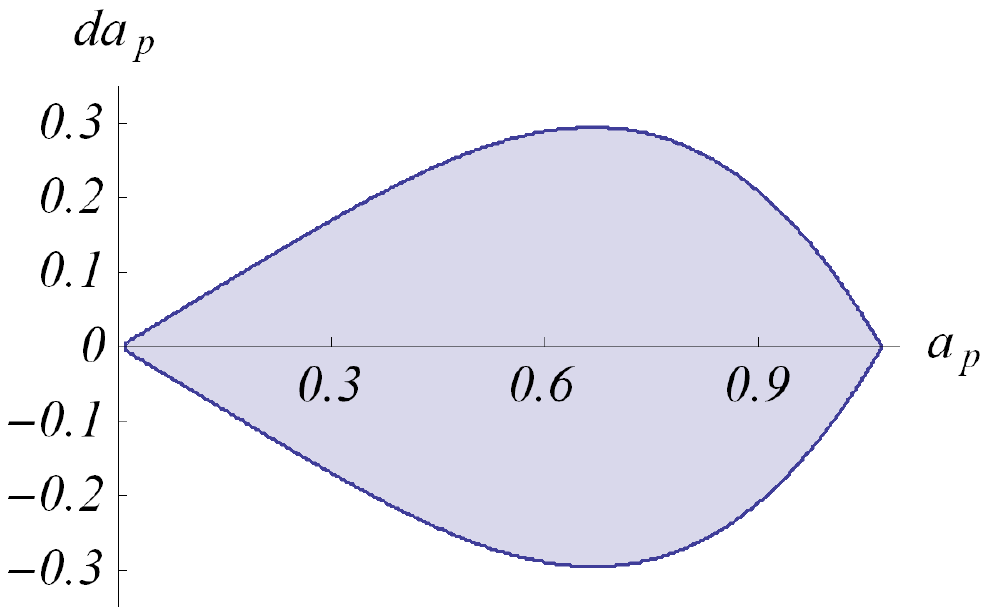}
  \caption{\label{Domainep}Admissible domain in the $(a_p,da_p)$ plane.}

\end{subfigure}
\begin{subfigure}{.5\textwidth}
  \centering
  \includegraphics[width=\linewidth]{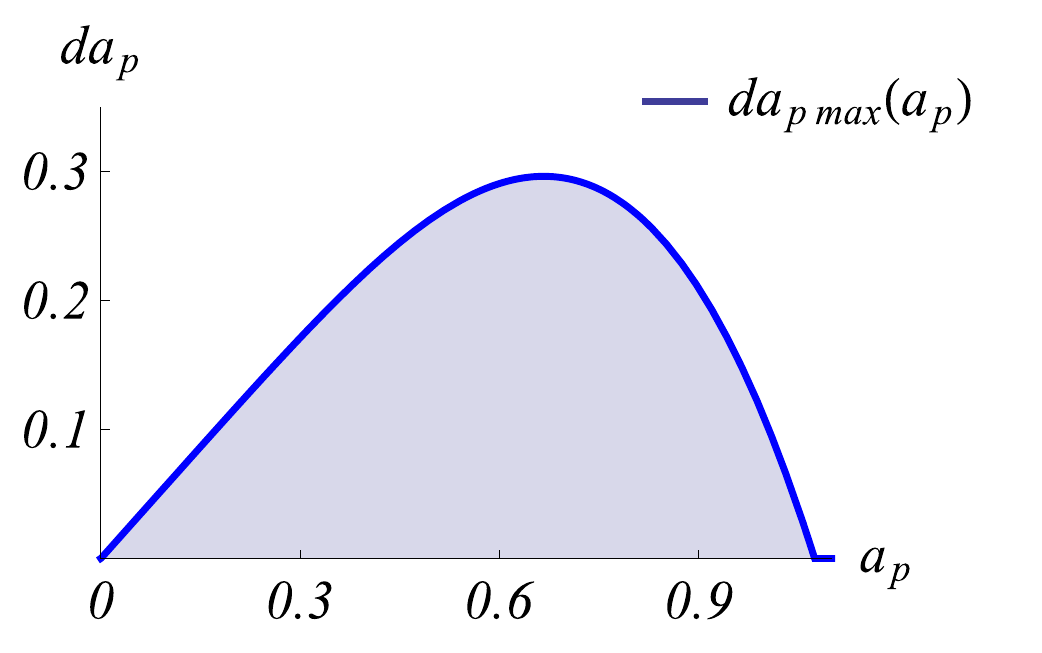}
  \caption{\label{dapmax}Plot of $da_{p,max}$ as a function of $a_p$.}

\end{subfigure}

\caption{\label{Domainesp}Admissible domain for IC at  $\frac{\pi}{2}$}
\end{figure}

Another possibility is to impose IC for $a$, $c$ at $\theta=\frac{\pi}{2}$. 
However, as already mentioned, regularity of the solution requires  $a(\frac{\pi}{2})=c(\frac{\pi}{2})$. We should therefore use  
$a_p:=a(\frac{\pi}{2})$ and $da_p:=a'(\frac{\pi}{2})$ instead 
as independent parameters; we can see this clearly by considering the 
analytic expansion of the solution at  $\theta=\frac{\pi}{2}$:
\begin{equation}\label{Dvptac}
	\begin{array}{r c l}
	a(\frac{\pi}{2}-\varepsilon)& =& a_p + da_p \varepsilon + (-a_p + a_p^5 + 4\frac{da_p^2}{a_p}) \frac{\varepsilon^2}{2} +\mathcal{O}(\varepsilon^3) \\
	c(\frac{\pi}{2}-\varepsilon)& =& a_p - a_p(1+a_p^4)\frac{\varepsilon^2}{2} +\mathcal{O}(\varepsilon^3) 
	\end{array}
~.
\end{equation}
However it is difficult to perform the numerical analysis with this type of IC. 
To circumvent this problem we will take as IC the values of $a$ and  $c$ evaluated at $\frac{\pi}{2}-\epsilon$, with $\epsilon$ of the order of $10^{-4}$, and we will use (\ref{Dvptac}) to express $a(\frac{\pi}{2}-\epsilon)$ and $c(\frac{\pi}{2}-\epsilon)$ in terms of $a_p$ and $da_p$.
 We can now plot the admissible domain in the $(a_p,da_p)$ plane, as shown in figure \ref{Domainep}. 
The domain is symmetric with respect to the $da_p=0$ axis. 
We can also plot the function $da_{p,max}$,
which gives the maximum value of $da_p$ for a given  $a_p$ (figure \ref{dapmax}).

\begin{figure}
\begin{subfigure}{.5\textwidth}
  \centering
  \includegraphics[width=\linewidth]{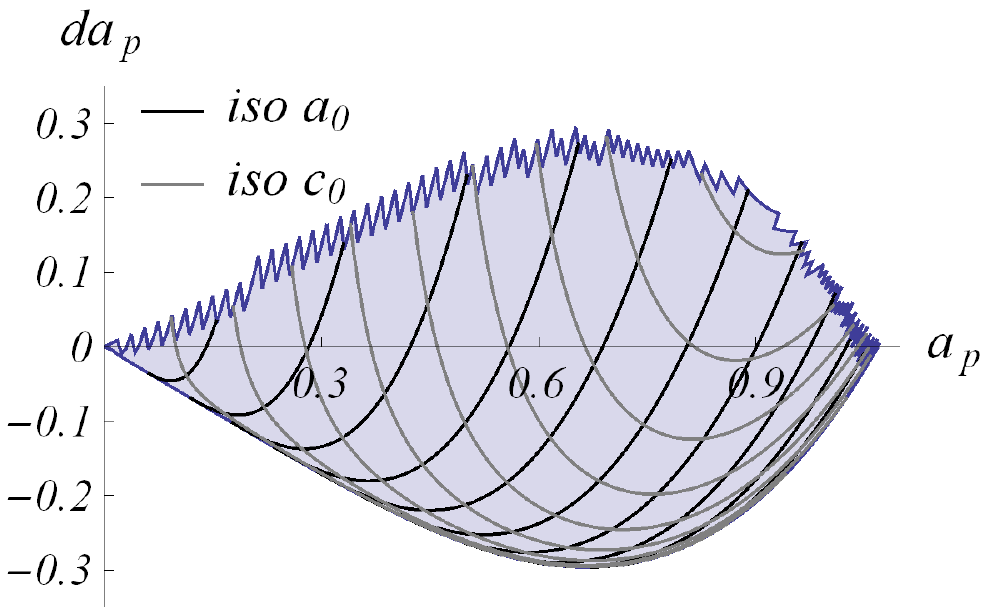}
  \caption{\label{0topi2} Admissible domain in the $(a_p,da_p)$ plane.}

\end{subfigure}
\begin{subfigure}{.5\textwidth}
  \centering
  \includegraphics[width=\linewidth]{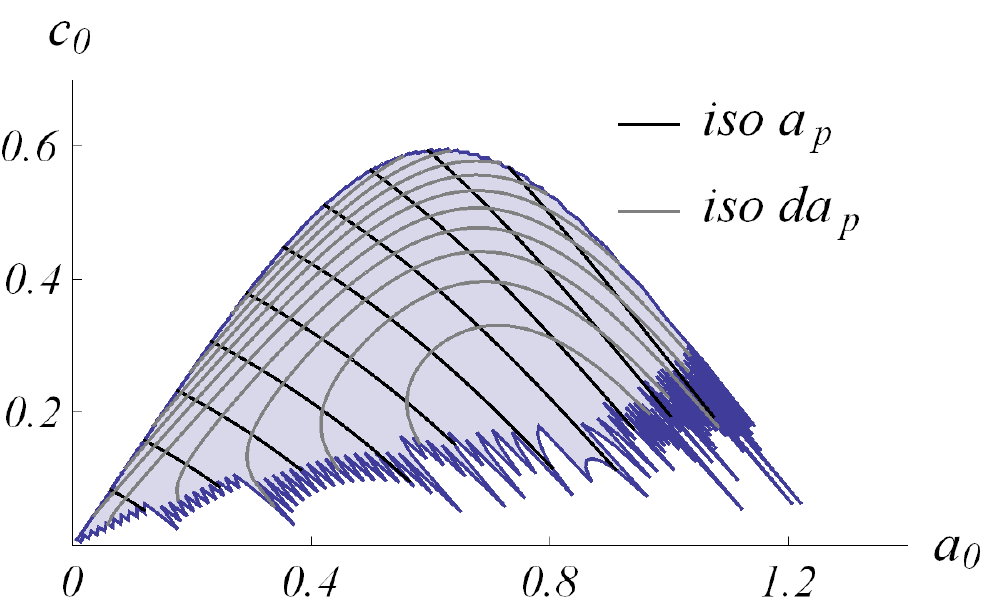}
  \caption{\label{pi1to0}  Admissible domain in the $(a_0,c_0)$ plane.}

\end{subfigure}

\caption{\label{correspondance2}Correspondance between the two types of IC: the lines in each domain are iso-IC.}
\end{figure}

It is also possible  to show explicitly the equivalence of the two different types of IC considered above. To that end 
we have worked out the correspondence between the two domains: 
in figure \ref{correspondance2} 
we have ploted the domain sweeped by the IC at $\theta=\frac{\pi}{2}$ 
as functions of the IC at $\theta=0$ and conversely. This also 
allows us to detect the areas which require increased precision depending 
on the method used: we see that using IC at $\theta=0$ favors 
small values of  $da_p$; conversely using initial conditions at 
$\theta=\frac{\pi}{2}$ favors high values of $a_0$.

\subsection{Eigenvalues}\label{sec:vp}

Let us now come to the eigenvalue problem \eqref{Sturm}. 
As discussed in section \ref{sec:sl} the transformation $\tilde{g}:=g/u$, with $u$ given in (\ref{udef}) transforms (\ref{Sturm}) into the eigenvalue problem 
(\ref{Sturmreg}). 
When supplemented with separated BC:
\eq{\spl{\label{se}
A_1&\tilde{g}(0)+A_2(\tilde{p}\tilde{g}')(0)=0, ~~ A_1^2+A_2^2\neq 0\\
B_1&\tilde{g}(\pi)~\!+B_2(\tilde{p}\tilde{g}')(\pi)=0, ~~ B_1^2+B_2^2\neq 0
~,}}
for some real numbers $A_1$, $A_2$, $B_1$, $B_2$, (\ref{Sturmreg}) becomes a well-posed, regular SL problem whose spectrum 
of eigenvalues:
\eq{\label{spec}\lambda\in\{0,~L^2M_1^2,~L^2M_2^2,\dots,~L^2M_n^2,\dots\}~.}
coincides with that of (\ref{Sturm}).

{}Furthermore let us consider the BC that must be imposed at $\theta=0,\pi$. According to the discussion of section \ref{sec:s2} these should be chosen so that 
(\ref{norm}), (\ref{intcond}) are satisfied. 
Taking into account the explicit form of the metric (\ref{3}), these turn out to be equivalent  to the conditions
\eq{\label{norm2}\int_{0}^{\pi}\d\theta qg^2<\infty~,}
and
\eq{\label{intcond2}\int_{0}^{\pi}\d\theta\left( pgg'\right)'=0~,}
respectively. 
On the other hand solving (\ref{Sturm}) perturbatively in the neighborhood of $\theta=0$ yields the following expansion:
\begin{equation}\label{Devg}
	g(\theta) =  c_{1}+c_{2}\ln\theta -\frac{\lambda c_{1}}{16} \theta^2 +\left( -\frac{c_{2}}{6}+\frac{c_{2}}{16} \lambda\left(1-\ln\theta \right)\right)\theta^2   +  \mathcal{O}(\theta^3\ln\theta)
~,
\end{equation}
for some arbitrary constants $c_1$, $c_2$, in accordance with the fact that 
$\theta=0$ is a regular singular point of (\ref{Sturm}). Taking into account 
that $p$, $q\sim\theta$ near $\theta=0$ we conclude that (\ref{norm2}), (\ref{intcond2}) are satisfied if and only if $c_2=0$ in the expansion (\ref{Devg}), i.e., $g(\theta=0)$ is finite. Furthermore, taking into account 
the definitions (\ref{coeffs}), (\ref{udef}), it is straightforward to see that the latter condition is equivalent to the vanishing of $\tilde{p}\tilde{g}'$ at $\theta=0$. 
A similar analysis in the neighborhood of $\theta=\pi$ yields the condition that $\tilde{p}\tilde{g}'$ should vanish at $\theta=\pi$ or, equivalently, that $g(\theta=\pi)$ is finite. 

In conclusion, we have shown 
that the appropriate BC for the regular SL problem (\ref{Sturmreg}) are given by (\ref{sep}) of section \ref{sec:sl}; they are thus a special case of the 
separated BC (\ref{se}) 
obtained by setting $A_1=B_1=0$ therein. Moreover they are equivalent to the condition of finiteness of $g(\theta)$ at the endpoints $\theta=0,\pi$.

\begin{figure}
\begin{subfigure}{.5\textwidth}
  \centering
  \includegraphics[width=7.83cm]{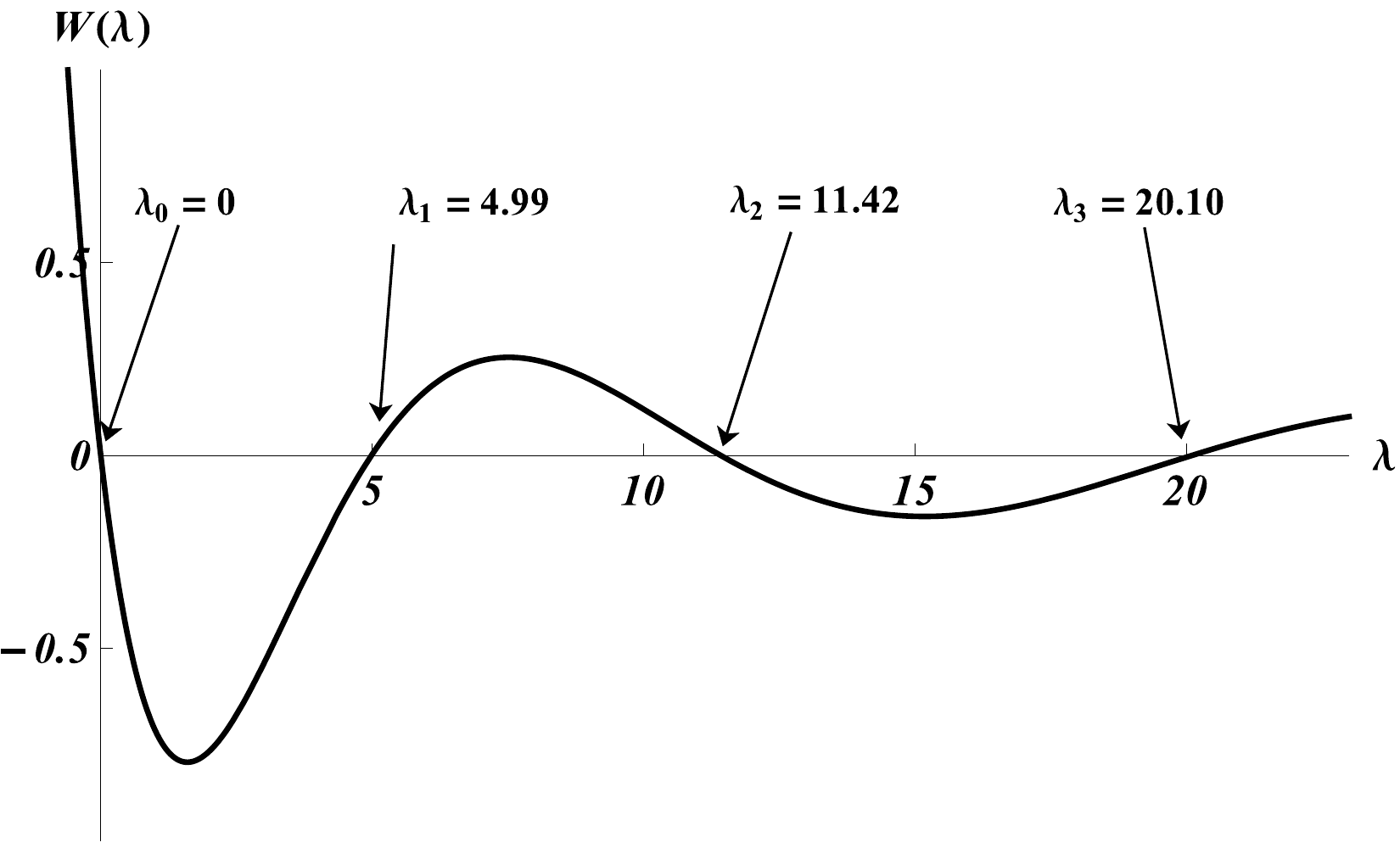}
  \caption{\label{Wronskienreg}Successive zeros of the Wronskian.}

\end{subfigure}
\begin{subfigure}{.5\textwidth}
  \centering
  \includegraphics[width=8.05cm,viewport=1 2 550 325,clip]{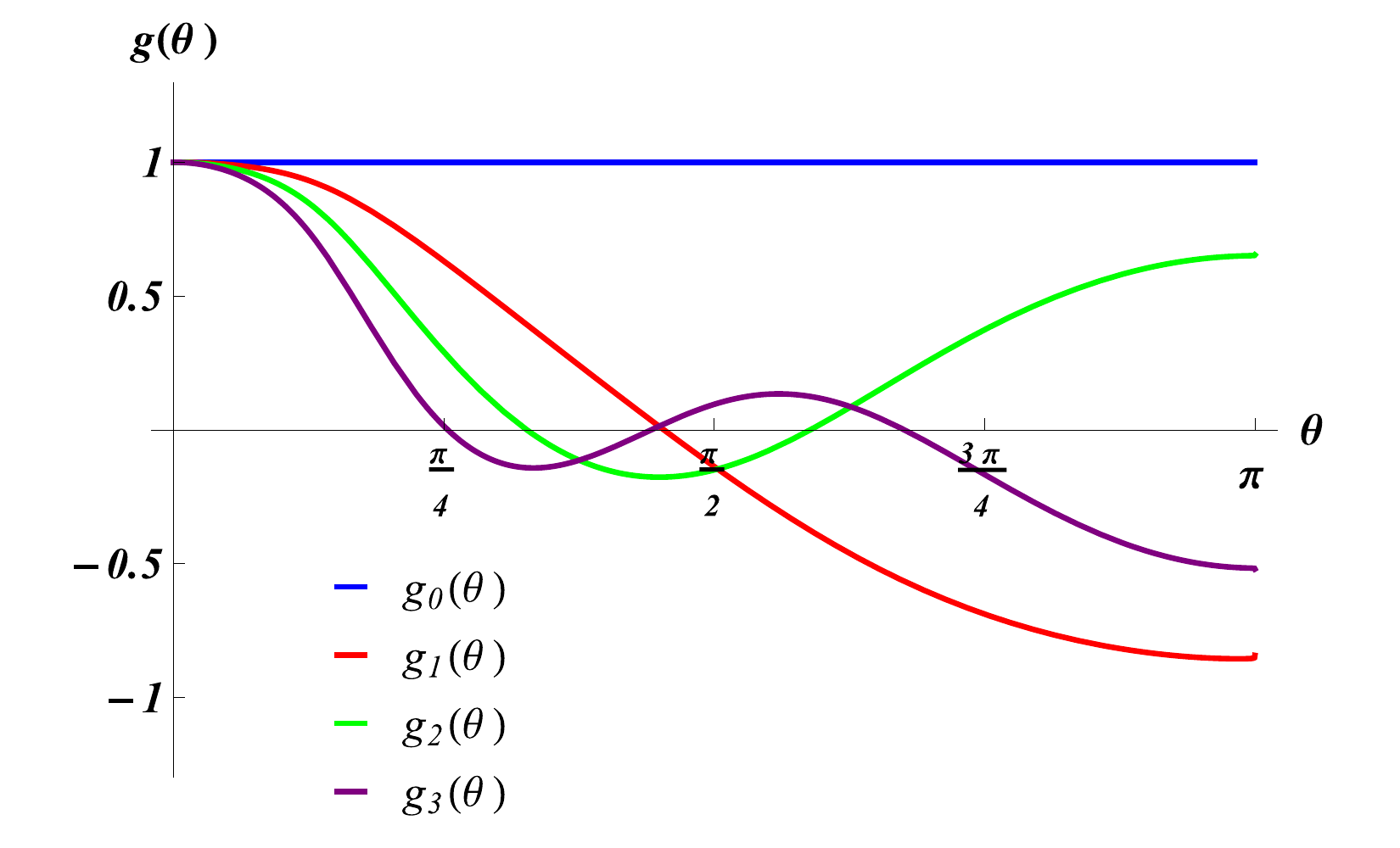}
  \caption{\label{Solsreg}The corresponding eigenfunctions.}

\end{subfigure}

\caption{\label{Spectrereg}Determination of the spectrum of the SL problem}
\end{figure}

\begin{figure}
\begin{subfigure}{.5\textwidth}
  \centering
  \includegraphics[width=\linewidth]{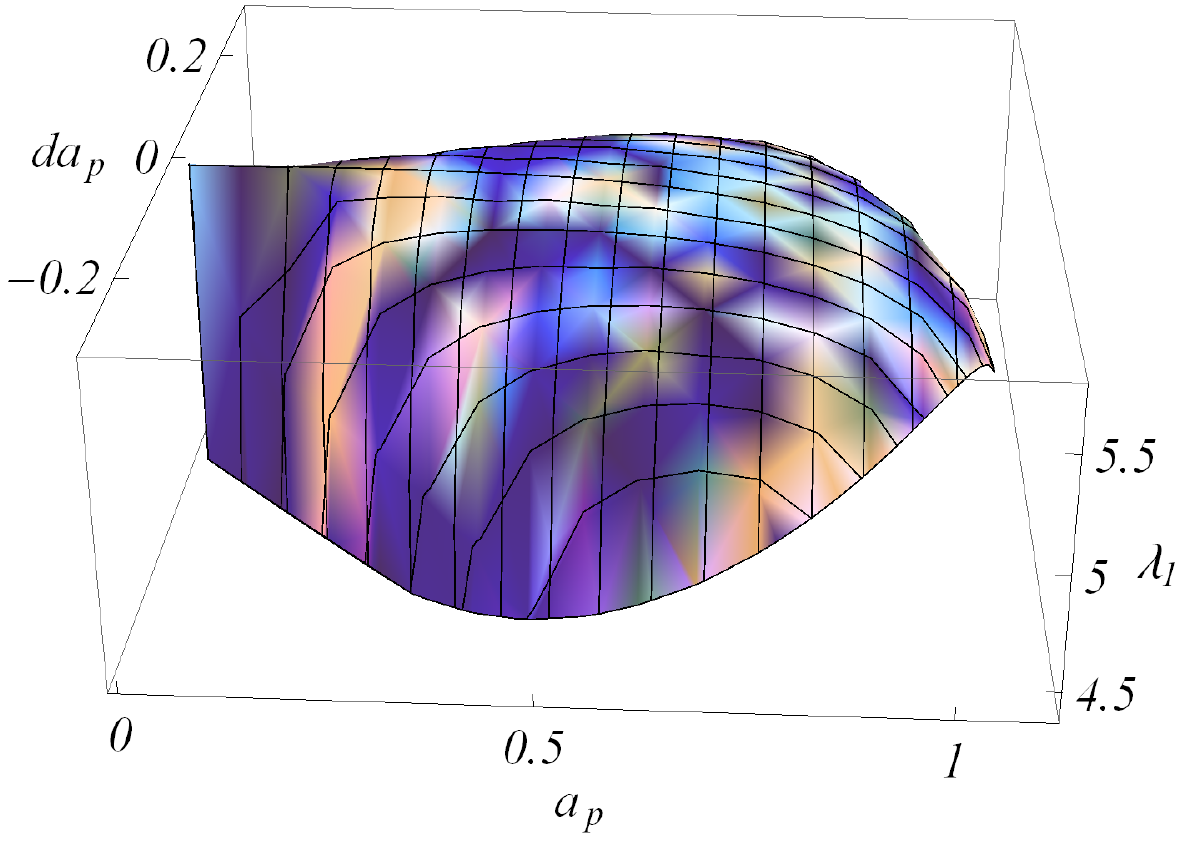}
  \caption{\label{Trace_vp2}Plot of $\lambda_1(a_p,da_p)$ over the admissible 
domain.}
\end{subfigure}
\begin{subfigure}{.5\textwidth}
  \centering
  \includegraphics[width=\linewidth]{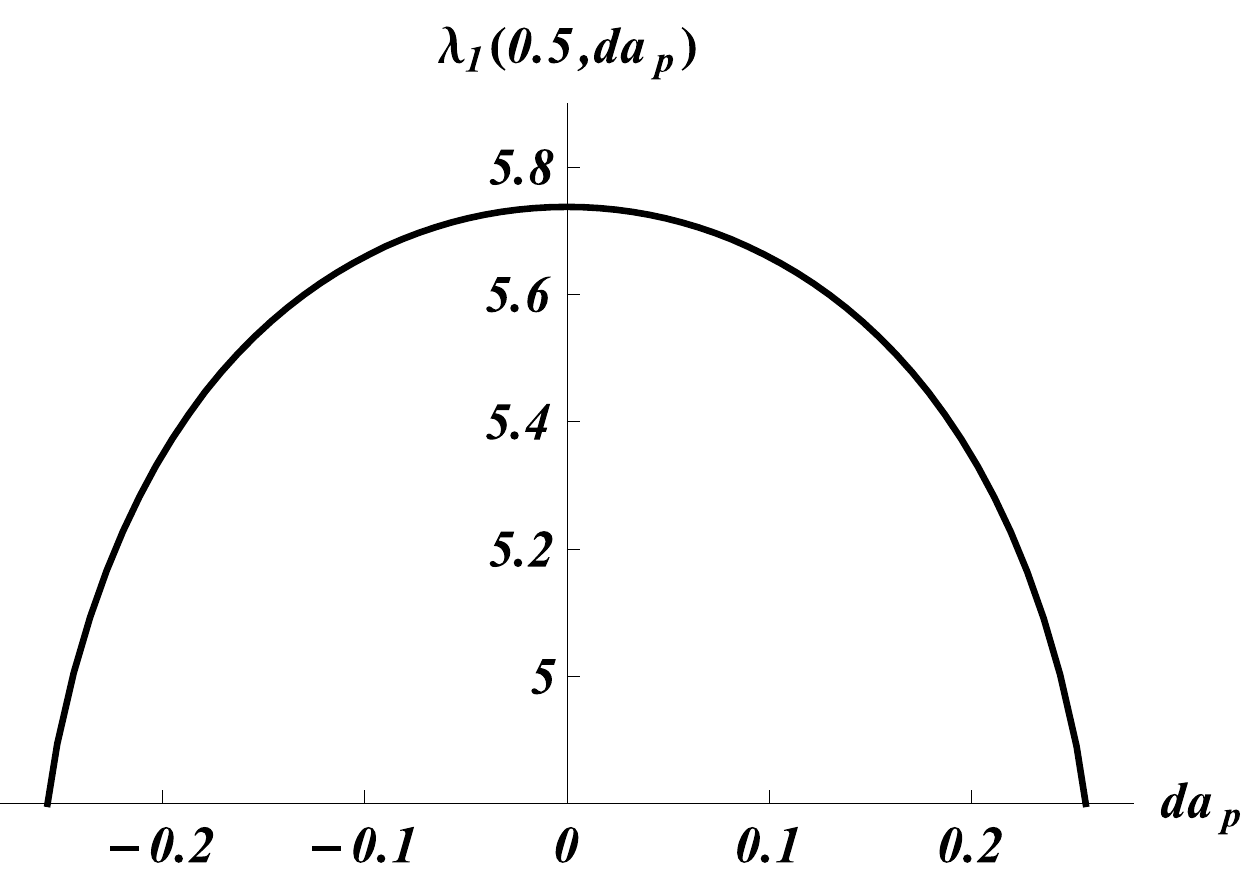}
  \caption{\label{Coupe_vp}The section $\lambda_1(a_p=0.5,da_p)$}

\end{subfigure}

\caption{\label{Surfacep}Determination of the KK scale.}
\end{figure}

\subsubsection*{Numerics}

As already mentioned, the eigenvalue problem (\ref{Sturm}), or the corresponding regular SL problem (\ref{Sturmreg}), when supplemented with the separated 
conditions (\ref{sep}) only has solutions for discrete values of $\lambda$, cf.,  (\ref{spec}). This can be seen as follows: for any  given $\lambda$ we can find a solution $\tilde{g}_{(\lambda)1}$ (unique up to normalization) of the differential equation (\ref{Sturmreg}) such that $\tilde{g}_{(\lambda)1}$ also satisfies the boundary condition (\ref{sep}) at $\theta=0$. Similarly, for the same $\lambda$ there exists a solution $\tilde{g}_{(\lambda)2}$ (unique up to an overall scale) of (\ref{Sturmreg}) such that $\tilde{g}_{(\lambda)2}$ satisfies the boundary condition (\ref{sep}) at $\theta=\pi$. For generic $\lambda$ these two solutions will be linearly independent unless their Wronskian vanishes:
\begin{align}
	W[\tilde{g}_{(\lambda)1},\tilde{g}_{(\lambda)2}] 
		:= \tilde{g}_{(\lambda)1} \tilde{g}_{(\lambda)2}'- \tilde{g}_{(\lambda)2} \tilde{g}_{(\lambda)1}'=0~,
\end{align}
in which case it is identically zero for all $\theta\in[0,\pi]$. 
The idea then of the numerical method for determining the spectrum of $\lambda$, which goes back to the work of Hartree \cite{hartree}, is to compute the Wronskian for a fixed $\theta\in[0,\pi]$ (we have chosen $\theta=\frac{\pi}{2}$ in our analysis) and to plot it as a function of $\lambda$. The values of $\lambda$ for which the Wronskian vanishes are the 
eigenvalues in the discrete spectrum (\ref{spec}) for which there exists a solution to the SL problem with separated BC. For example, in figure \ref{Wronskienreg} we have ploted 
the Wronskian as a function of $\lambda$; the first four zeros corresponding to the eigenvalues $\lambda_n$, $n=0,\dots,3$ are indicated explicitly; the coefficients $p(\theta)$, $q(\theta)$ of the SL problem (\ref{Sturm}) have been evaluated numerically for $a_0=1$, $c_0=0.1$. 
The corresponding eigenfunctions $g_n(\theta)$, $n=0,\dots,3$, are ploted in figure \ref{Solsreg}.

\begin{figure}
\begin{center}

\includegraphics[width=.5\textwidth]{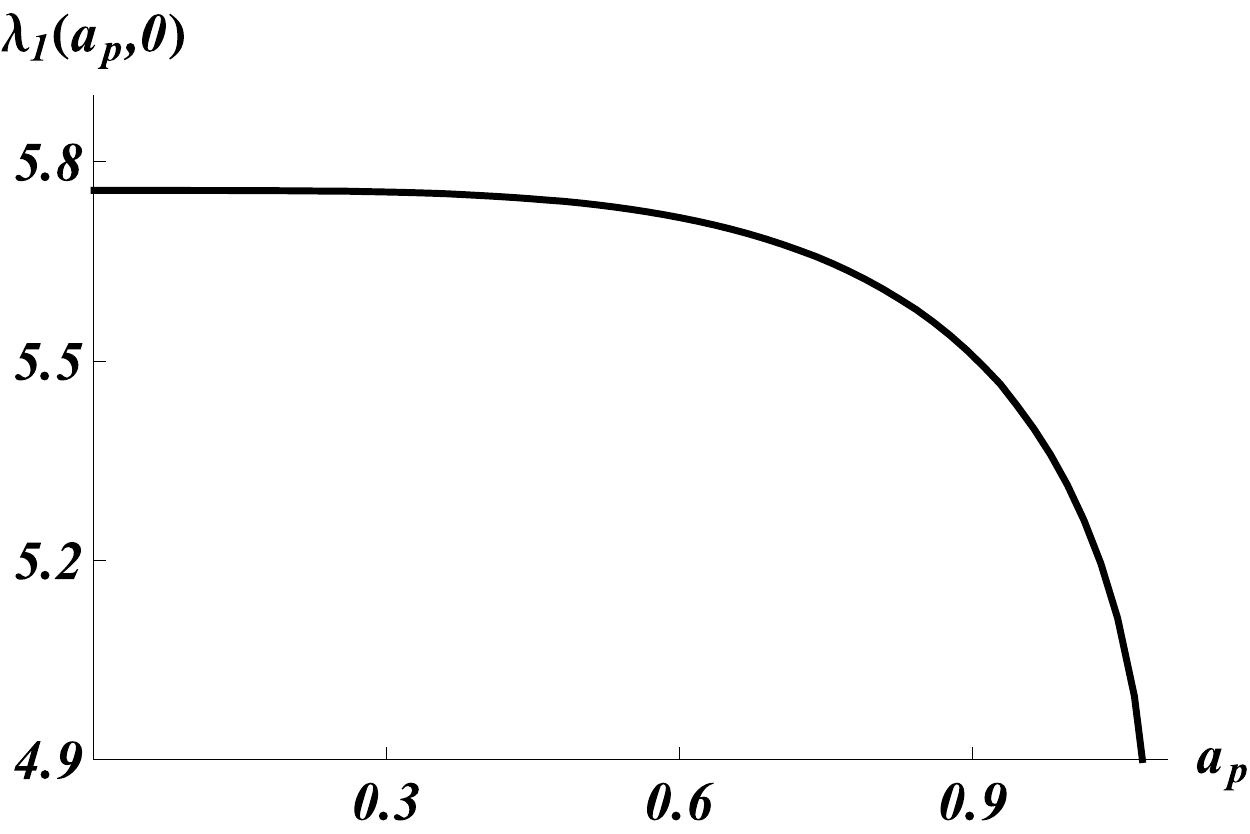}

\caption{\label{Coupe_da_p}Plot of $\lambda_1(a_p,da_p=0)$}

\end{center}
\end{figure}

In our case the goal is to determine the first nonvanishing eigenvalue $\lambda_1$ (which sets the KK scale) of the SL 
problem with separated BC (\ref{sep}). Since the coefficients $p(\theta)$, $q(\theta)$ of the SL problem (\ref{Sturm}) depend on the IC, $(a_0,c_0)$ 
or $(a_p,da_p)$, $\lambda_1$ can be thought of as a function of these IC. 
Applying the numerical method described above while varying the IC we can  
obtain $\lambda_1$ as a function of IC $\in D$ over the entire admissible domain. It is in fact more convenient for our purposes to use the IC  $(a_p,da_p)$ at $\theta=\frac{\pi}{2}$. We thus obtain the surface $\lambda_1(a_p,da_p)$ 
shown in figure  \ref{Trace_vp2}.
 The surface is symmetric with respect to the $da_p=0$ axis, so that 
the maximum of $\lambda_1$ is reached at $da_p=0$. This can be seen clearly in  the section shown in figure \ref{Coupe_vp}. 
In order to determine the maximum of $\lambda_1$ we should therefore focus on the section $da_p=0$. Its plot is depicted in figure \ref{Coupe_da_p}. 
We thus arrive at the result $\lambda_1 \leq 5.76$, which is equivalent to  (\ref{bound}) of the introduction.


\section{Conclusions}\label{sec:5}

We have performed a numerical study of the four-dimensional spin-2 KK spectrum in the supersymmetric AdS$_4$ vacua of \cite{lt3}. In particular  
we have shown that these vacua do not exhibit scale separation between the radius of AdS$_4$ and that of the compactification manifold $S^2(\mathcal{B}_4)$. It thus still remains an open challenge to construct pure-flux supersymmetric AdS supergravity solutions with scale separation, or to 
prove that such solutions are generally impossible.

The fact that the compactification geometry in our case is not known analytically (it is specified in terms of a coupled system of two first-order ODE's which has no known  analytic solution) means that the KK masses are 
determined from the eigenvalues of a SL problem which itself can only be defined numerically. In other words we had to solve a numerical eigenvalue problem on top of another numerical problem.

To our knowledge the methods employed in the present paper, albeit rather common in dealing with quantum-mechanical systems (this stems from the classic result that the one-dimensional Schr\"{o}dinger equation can be put in the form of a SL eigenvalue problem), have never been used before  in the context of Kaluza-Klein supergravity. They could be employed to treat similar problems where the compactification geometry is not known analytically, as in, e.g., \cite{Tomasiello:2010zz}. 
Our methods could equally well be used to  determine the KK spectrum on compactification spaces which are known analytically, but on which harmonic analysis may be cumbersome. 
Indeed the explicit analysis of the mass spectrum in KK supergravity is typically limited to compactifications on spheres and homogenous spaces \cite{Duff:1986hr}; this list of spaces can be considerably extended if one is willing to use a numerical approach.

In the case of compactifications which are known to exhibit scale separation,\footnote{Going beyond classical supergravity, there are several ways one can achieve scale separation.  For some recent results on supersymmetric vacua with scale separation and a survey of older work on the subject, see \cite{sa,sb}.} 
the difficulty in identifying the light KK masses (in order to determine the low-energy effective action) is sometimes dealt with by using consistent truncations that reduce the infinite number of KK modes to a finite set. However the resulting lower-dimensional theories are not in general guaranteed to be low-energy effective actions, i.e., to capture the physics of all the light modes. It would be interesting to examine whether the methods of the present paper may be used to shed light into this problem.


\begin{thebibliography}{99}

\bibitem{lt3}
  D.~L\"{u}st and D.~Tsimpis,
  ``New supersymmetric AdS(4) type II vacua,''
  JHEP {\bf 0909} (2009) 098
  [arXiv:0906.2561 [hep-th]].


\bibitem{Tsimpis:2012tu}
  D.~Tsimpis,
  ``Supersymmetric AdS vacua and separation of scales,''
  JHEP {\bf 1208} (2012) 142
  [arXiv:1206.5900 [hep-th]].








\bibitem{gm}
  J.~P.~Gauntlett, D.~Martelli, J.~F.~Sparks and D.~Waldram,
  ``A new infinite class of Sasaki-Einstein manifolds,''
  \atmp{8}{2006}{987} 
  [arXiv:\hepth{0403038}].


\bibitem{ms2}
  D.~Martelli and J.~Sparks,
  ``Notes on toric Sasaki-Einstein seven-manifolds and AdS$_4$/CFT$_3$,''
  JHEP {\bf 0811} (2008) 016
  [\arXividhepth{0808.0904}].


\bibitem{pz}
  M.~Petrini and A.~Zaffaroni,
  ``N=2 solutions of massive type IIA and their Chern-Simons duals,''
  \arXividhepth{0904.4915}.





\bibitem{Bachas:2011xa}
  C.~Bachas and J.~Estes,
  ``Spin-2 spectrum of defect theories,''
  JHEP {\bf 1106} (2011) 005
  [arXiv:1103.2800 [hep-th]].

\bibitem{Csaki:2000fc}
  C.~Csaki, J.~Erlich, T.~J.~Hollowood and Y.~Shirman,
  ``Universal aspects of gravity localized on thick branes,''
  Nucl.\ Phys.\ B {\bf 581} (2000) 309
  [hep-th/0001033].





\bibitem{Duff:1986hr}
  M.~J.~Duff, B.~E.~W.~Nilsson and C.~N.~Pope,
  ``Kaluza-Klein Supergravity,''
  Phys.\ Rept.\  {\bf 130} (1986) 1.



\bibitem{zbook}
  A.~Zettl,  
  ``Sturm-Liouville Theory,''
  Amercian Mathematical Society, Providence, RI, 2005.



\bibitem{math} 
 Wolfram Research, Inc., Mathematica, Version 10.0, Champaign, IL, 2014.



\bibitem{hartree}
D.R.~Hartree, ``The calculation of atomic structures,'' Wiley, 
New York, NY, 1957.


\bibitem{Tomasiello:2010zz}
  A.~Tomasiello and A.~Zaffaroni,
  ``Parameter spaces of massive IIA solutions,''
  JHEP {\bf 1104} (2011) 067
  [arXiv:1010.4648 [hep-th]].


\bibitem{sa}
  T.~Maxfield, J.~McOrist, D.~Robbins and S.~Sethi,
  ``New Examples of Flux Vacua,''
  JHEP {\bf 1312} (2013) 032
  [arXiv:1309.2577 [hep-th]].


\bibitem{sb}
  J.~McOrist and S.~Sethi,
  ``M-theory and Type IIA Flux Compactifications,''
  JHEP {\bf 1212} (2012) 122
  [arXiv:1208.0261 [hep-th]].

\end{thebibliography}
\end{document}